\documentclass{ifacconf} 

\usepackage{amssymb}
\usepackage{amsfonts}                                
\usepackage[latin1] {inputenc}
\usepackage{enumerate}
\usepackage{mathrsfs}
\usepackage[round,authoryear,longnamesfirst]{natbib}
\usepackage{psfrag}
\usepackage{graphicx}          
\usepackage{color}      

\def\qedp{\hspace*{\fill}~{\tiny $\blacksquare$}}
\def\qed{\relax\ifmmode\hskip2em \Box\else\unskip\nobreak\hskip1em $\Box$\fi}

\newtheorem{theorem}{Theorem}

\newtheorem{itdefinition}{Definition}
\newtheorem{itproposition}{Proposition}
\newtheorem{itresult}{Result}
\newtheorem{itremark}{Remark}
\newtheorem{itassumption}{Assumption}
\newtheorem{itcorollary}{Corollary}
\newtheorem{itexample}{Example}
\newenvironment{definition}{\begin{itdefinition}\rm}{\end{itdefinition}}
\newenvironment{proposition}{\begin{itproposition}\rm}{\end{itproposition}}
\newenvironment{remark}{\begin{itremark}\rm}{\end{itremark}}
\newenvironment{assumption}{\begin{itassumption}\rm}{\end{itassumption}}

\begin{document}

\begin{frontmatter}

\title{Resilient Control under Denial-of-Service} 

 
\author[CastleRock]{C. De Persis}   
\author[CastleRock]{P. Tesi} 
                  
\address[CastleRock]{ITM, Faculty of Mathematics and Natural Sciences, University of Groningen, 9747 AG Groningen, The Netherlands \\
{\tt \{c.de.persis, p.tesi\}@rug.nl}  }             
         

\begin{abstract}
We investigate resilient control strategies for linear systems under Denial-of-Service (DoS) attacks. 
By DoS attacks we mean interruptions of communication on measurement (sensor-to-controller)  
and/or  control (controller-to-actuator) channels carried out by an intelligent adversary.  
We characterize the duration of these interruptions under which stability of the closed-loop system is preserved. 
The resilient nature of the control descends from its ability to adapt the sampling rate to the occurrence of the DoS.
\end{abstract}

\end{frontmatter}

\section{Introduction} \label{sec:introduction}

In recent years there has been a growing interest concerning  feedback control systems 
that are implemented over communication networks. These networks impose that measurements 
are acquired at discrete times, transmitted and received by the controller. The latter processes the 
received information and computes the control signal. This can in turn  be sampled and transmitted 
to the actuators. Common limitations on these signals that travel over a network are quantization, delays and loss of information. 
Due to the limited bandwidth of the communication channel, as well as possible constraints on the 
available computational power, much research has been devoted to reduce the use of the communication 
line, by designing the sampling sequence based on current status of the process to control. 
This has given raise to a very active line of research in
the context of \emph{event/self-triggering} control; see \cite{Heemels} for a recent 
comprehensive overview of the topic. 

In the literature, several aspects of event/self-triggering control have been investigated, 
including output-feedback (\cite{donkers.heemels.ifac11}), robustness against additive 
disturbances (\cite{mazo2010iss}), large-scale systems (\cite{wang2011, {CDP-RS-FW:AUT13}}) 
and distributed coordinated control (\cite{GS-DVD-KHJ:13, CDP:PF:TAC13}), to name a few.
On the other hand, an aspect of primary importance for which less results are available is the robustness of such schemes
against malicious attacks. 

Attacks to computer networks have become ever more prevalent over the last few years. 
In this respect, one of the (if not the) most common type of attack is 
the so-called \emph{Denial-of-Service} (DoS, for short); see \cite{Lowe}
for an introduction to the topic.
It consists in a radio interference signal (also known as \emph{jamming} signal)  
which is primarily intended to affect the timeliness of the information exchange,
\emph{i.e.} to cause packet losses. While 
networked control formulations have previously considered sensor/control packet losses (\cite{Schenato}),
dealing with DoS phenomena requires fundamentally different analysis tools. 
In fact, in contrast with classical networked control systems
where packet losses can be reasonably modeled as random events,
assuming a stochastic characterization of the DoS attacks would be inherently limiting
in that it would fail to capture the malicious and intelligent nature of an attacker. 

Prompted by these considerations, this paper discusses the 
problem of controlling networked systems subject to DoS attacks,
whose underlying strategy is \emph{unknown}. 
More specifically, we consider a classical \emph{sampled-data} control system consisting of a  
continuous-time linear process in feedback loop with a digital controller.
An attacker, according to some unknown strategy, 
can interrupt both sensor and control communication channels.
Under such circumstances, the process evolves under out-of-date control.
Within this context, we address the question of designing control update 
rules that are capable
 of ensuring closed-loop stability  
in the event that DoS does not occur too frequently
\footnote{Clearly, in the case of a system that is open loop unstable, 
a DoS that never allows the transmission of feedback information would easily lead to instability.}.
In this respect, the main contribution of this paper is to show that 
suitable control update rules do exist whenever the ratio between the
``active'' and ``sleeping'' periods of jamming is small enough on the average.
This somehow reminds of stability problems for systems that switch between stable and unstable modes; 
see e.g.~\cite{zhai.et.al}. In our paper, however, the peculiarity of the problem under study leads to a different analysis and results.  
We also show that the results here introduced
are flexible enough so as to allow the designer to choose 
from several implementation options that can be used 
to trade-off performance vs. communication
resources. Although these solutions originate from
fundamentally different approaches, they exhibit
the common feature of \emph{resilience}, by which
we mean the possibility to adapt the sampling rate to the DoS occurrence.

Previous contributions to this research line have been reported in
\cite{sastry,basar}. In these papers, however, the framework is substantially 
different. They consider a pure discrete-time setting in which 
the goal is to find optimal control and attack strategies 
assuming a maximum number of jamming actions over a prescribed (finite) control horizon.
Here, we do not formulate the problem as an optimal control design problem. 
The controller can be designed according to any suitable design method,
robustness and resilience against DoS attacks being achieved thanks 
to the design of the control update rule.
Perhaps, the closest references to our research is \cite{HSF-SM:13-ijrnc}.
In that paper, the authors consider a situation in which the 
attack strategy is known to be \emph{periodic}, though of 
unknown period and duration.
The goal is then to identify period and duration of the jamming activity
so as to determine the time-intervals where
communication is possible.
Their framework should be therefore looked at as complementary 
more than alternative to the present one, 
since the former deals with cases where one can adjust the control updates so that 
they never fall into the jamming activity periods.
Such a feature is conceptually impossible to achieve
in scenarios such as the one considered in this paper, 
where the jamming strategy is neither known 
nor prefixed (the attacker can modify on-line the attack strategy). 

The remainder of this paper is organized as follows. In Section \ref{sec.problem} 
we introduce the class of DoS signals which are of interest in this paper and formulate 
the control problem. The main result  with a characterization of the class 
of DoS signals under which stability is preserved is given in Section \ref{sec.main}.  
Section \ref{sec:switch} discusses implementation issues and presents a 
number of resilient control strategies. 
In Appendix A, we report an alternative Lyapunov-based treatment  of the main result. 

Proofs of the main result and the
auxiliary lemmas are reported in Appendix B.

\section{Framework and problem overview}\label{sec.problem}

\begin{figure}[tb]
\psfrag{x}{{\tiny $x$}}
\psfrag{u}{{\tiny $u$}}
\includegraphics[width=0.45 \textwidth]{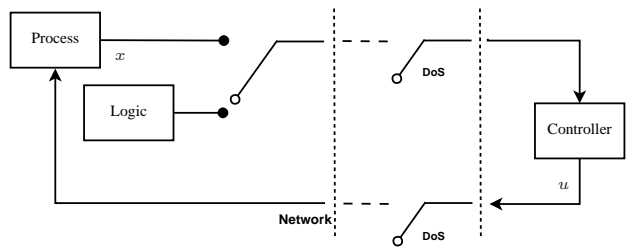} 
\linespread{1}\caption{Block diagram of the closed-loop system 
under DoS on the communication channels.} \label{fig:BasicScheme}
\end{figure}

The framework of interest can be schematically 
represented as in Figure \ref{fig:BasicScheme}. Specifically, we consider
a remote plant-operator setup, in which
the process to be controlled 
is described by the differential equation
\begin{equation} \label{system}
\dot x(t) = A x(t) + B u(t)
\end{equation}
where $t \in \mathbb R_{\geq 0}$; $x \in \mathbb R^{n_x}$ is the state
and $u \in \mathbb R^{n_u}$ is the control input; $A$ and $B$ are 
matrices of appropriate size. We assume that a state-feedback
matrix $K$ has been designed rendering the closed-loop system 
stable in the sense of Lyapunov, \emph{i.e.} such that all
the eigenvalues of $A+BK$ have negative real part.

The control action is implemented 
via a \emph{sample-and-hold} device. Let 
$\{t_k\}$, $k \in \mathbb N$, $t_0:=0$ by convention, represent the sequence 
of time instants at which it is desired to update the control action.
At the present stage, for simplicity of exposition, we shall simply refer to 
the ``Logic'' block as the device responsible for generating $\{t_k\}$.
Accordingly, whatever the logic underlying this block, in the ideal situation 
where data can be sent and received at any desired instant of time, 
the control input applied to the process is given by
\begin{equation}  \label{ideal_control}
u_{{\rm ideal}}(t) = K\,x(t_k), \quad \forall \, t \in [t_k,t_{k+1}[ 
\end{equation}

We shall refer to \emph{Denial-of-Service} (DoS, for short) 
as the phenomenon that may prevent (\ref{ideal_control})
from being applied over certain time periods. In principle,
such a phenomenon can affect both control (controller-to-actuator)  
and measurement (sensor-to-controller)  
channels. In this paper, we consider the
case of a DoS \emph{simultaneously} affecting both 
control and measurement channels. This amounts to assuming 
that, in the presence of DoS, data can be neither sent nor received.
To make this concept precise, let 
$\{h_n\}$, $n \in \mathbb N$, $h_0 \geq 0$, represent the sequence 
of DoS \emph{positive edge-triggering} 
\footnote{Borrowing the terminology from digital applications, 
by ``positive edge-triggering'' we mean
the time instants at which the DoS exhibits 
a transition from, say, zero (communication is possible) to, say, one 
(communication is interrupted). This, along with $\tau_n >0$, implies
that $h_{n+1}>h_{n}$ for all $n \in \mathbb N$.}
and
\begin{equation}  \label{DoS_intervals}
H_n = [h_n,h_n+\tau_n[  
\end{equation}
with $\tau_n >0$, the corresponding DoS $n$-th time-interval.
In this respect, we shall assume that over each $H_n$  
the actuator generates an input that is 
based on the most recently received control signal. 
Accordingly, let
\begin{eqnarray}  \label{last_DoS}
n(t)= \left\{ 
\begin{array}{ll} -1, & \qquad \textrm{if} \,\, t<h_0  \\ \\
\sup \, \{ \, n \in \mathbb N \, | \,\, h_n < t \,   \}, & \qquad \textrm{otherwise}
\end{array}
\right.
\end{eqnarray}
denote the last (up to the current time) DoS positive edge-triggering,
and let
\begin{eqnarray}  \label{last_succesful_control_update}
k(t)= \left\{ 
\begin{array}{ll} 
-1, & \textrm{if} \,\, h_0 = 0  \\  
& \textrm{and} \,\, t \in H_0  \\  \\
\begin{array}{l}
\sup \, \left\{ \, k \in \mathbb N\, | \,\, t_k < t; \, \right. \\  
\qquad \qquad  \quad \,\,\,\, \left. t_k \notin \bigcup_{n=0}^{n(t)} H_n  \,   \right\}, 
\end{array}
& \textrm{otherwise}
\end{array}
\right.
\end{eqnarray}
denote the last (up to the current time) 
successful control update. Then, 
the actual control input applied to the process 
is given by
\begin{equation}  \label{actual_control}
u(t) = K\,x(t_{k(t)}) 
\end{equation}
The possible presence of DoS also raises 
the question of assigning a value for control input in case
$h_0=0$, \emph{i.e.} when communication is not possible 
at the process start-up. In this respect, in case $h_0=0$,
we shall assume that $u(0) = 0$ and let $x(t_{-1}) := 0$
for notational consistency.

\begin{remark} It would be possible to consider an alternative framework 
in which the input to the process is always zero during DoS intervals \citep{basar}. 
As will become clear in the following, the latter can be regarded
as a simplified variant of the framework in which, upon DoS,
the process input is based on the most recently received control signal. 
\hfill $\Box$
\end{remark}

\subsection{Problem overview}

To begin with, it is convenient to introduce the following 
definition.

\begin{definition} \label{def:GES}
Consider the control system $\Sigma$ composed of 
(\ref{system}) under a state-feedback control as in (\ref{actual_control}). 
$\Sigma$ is said to be \emph{globally exponentially stable} (GES) 
if there exist $\alpha,\beta \in \mathbb R_{>0}$ such that
\begin{equation} \label{GES}
\| x(t) \| \leq \alpha e^{-\beta t} \| x(0) \| 
\end{equation}
for all $t \in \mathbb R_{\geq 0}$ and for all $x(0) \in \mathbb R^{n_x}$,
where $\|\cdot\|$ stands for Euclidean norm. 
\hfill $\Box$
\end{definition}

Various approaches have been considered 
assuring GES to the control system in the absence of DoS;
\emph{e.g.}, see \cite{Heemels} for recent results and
a discussion on questions related to periodic vs aperiodic implementations. 
A natural question then arises on whether mechanisms do exist
that are capable of preserving GES for certain DoS signals.

In this respect, some preliminary considerations are in order.
Whatever the rule generating the $\{t_k\}$-sequence, ultimate goal of the ``Logic'' block
is to update the control action frequently enough so that stability
is not destroyed. While in principle this is always possible in the absence of DoS 
\footnote{In the absence of DoS, (\ref{actual_control}) 
reduces to (\ref{ideal_control}). Then, assuming sufficient computational resources,
stability can always be ensured by choosing $\{t_k\}$ in such a way that
the inter-sampling times are small enough.}, the same conclusions do not hold 
if DoS is allowed to be arbitrary.
For instance, for open-loop unstable systems, one immediately sees that if 
$\tau_0 = \infty$ then 
stability is lost irrespective of how $\{t_k\}$ is chosen. 
These points motivate the following restriction 
on the admissible DoS signals considered throughout the paper.

Given a sequence $\{h_n\}$, let
\begin{eqnarray}  \label{DoS_intervals_union}
\Xi(t) = \left\{ \, \bigcup_{n=0}^{n(t)-1} H_n \; \right\} \, \bigcup \, \left[h_{n(t)}, \min \{h_{n(t)}+\tau_{n(t)}; \,t\} \right] \nonumber \\
\end{eqnarray}
denote the sum of DoS intervals up to the current time,
and, given an interval $I$, let $|I|$ denote its length.

\begin{assumption} \label{ass:DoS_slow}
The DoS sequence $\{h_n\}$, $n \in \mathbb N$,  is such that 
$\inf_{n \in \mathbb N} \tau_n >0$. Moreover,
there exist constants $\kappa \in \mathbb R_{\geq0}$ and $\tau \in \mathbb R_{>0}$ such that
\begin{equation}  \label{DoS_slow}
|\Xi(t)| \, \leq \, \kappa + \frac{t}{\tau}
\end{equation}
for all $t \in \mathbb R_{\geq 0}$. 
\hfill $\Box$
\end{assumption}

\begin{remark} \label{rem:knowledge}
Condition $\inf_{n \in \mathbb N} \tau_n >0$ ensures that $\{h_n\}$ is 
sufficiently ``regular''. In particular, it implies that $\{h_n\}$ is \emph{non-Zeno} and that
an infinite number of DoS intervals does always have strictly positive Lebesgue measure. 
Inequality (\ref{DoS_slow}) expresses the property that the DoS satisfies 
a \emph{slow-on-the-average} type condition, as introduced by \cite{Hespanha} for hybrid systems analysis.
In the present context, the rationale behind (\ref{DoS_slow}) is that if $\kappa=0$
then the average time interval of DoS is at least $\tau$. On the other hand, $\kappa>0$
allows for consideration of DoS at the process start-up, \emph{i.e.} when $h_0=0$.
\hfill $\Box$
\end{remark}

\begin{remark} \label{rem:knowledge}
The considered framework differs from classical networked systems
where information loss can be, for instance, assumed to follow
some suitable stochastic distribution; \emph{e.g.}, see \cite{Schenato}. 
In fact, concepts such as DoS, cyber-physical security/resilience 
are mainly oriented towards scenarios where communication loss
can result from threats of a malicious nature. 
This is the primary motivation 
for not assuming $\{h_n\}$ to follow a specific distribution or pattern.
\hfill $\Box$
\end{remark}

\section{Main results}\label{sec.main} 

In this section, a simple control update rule is considered,
which is capable of preserving GES for any DoS signal 
satisfying Assumption \ref{ass:DoS_slow} with 
$\tau$ sufficiently large. A discussion on the results
along with implementation aspects is deferred to the next section.

Let
\begin{equation} \label{error}
e(t) := x(t_{k(t)}) - x(t) 
\end{equation}
where $t \in \mathbb R_{\geq 0}$, represent the error between the 
value of the process state at the last successful control update 
and the value of the process state at the current time. 
Consistent with the comments made right after (\ref{actual_control}),
if $h_0=0$ then $e(t)=-x(t)$ for all $t \in H_0$.
The closed-loop system composed of 
(\ref{system}) and (\ref{actual_control})
can be therefore rewritten as
\begin{equation} \label{control_system}
\dot x(t) \, = \, \Phi x(t) + BK e(t) 
\end{equation}
where $\Phi:=A+BK$. Consider now the following 
control update rule
\begin{equation} \label{control_update_rule}
\|e(t)\| \, \leq \, \sigma  \|x(t)\|, \quad \forall \, t \notin \Xi(t) 
\end{equation}
where $\sigma \in \mathbb R_{> 0}$ is a free design parameter. 
As shown hereafter, such an update rule is capable of preserving 
GES for any DoS signal satisfying Assumption \ref{ass:DoS_slow} 
with $\tau$ sufficiently large.

Condition (\ref{control_update_rule})
was first introduced in \cite{Tabuada07} in the context of event-based
control. The difference here is that, due to the presence of 
DoS, one cannot enforce this condition for all  $t\geq 0$,
but only over those time-intervals where communication is indeed possible.

\begin{figure}[tb]
\psfrag{t}{{\tiny $t$}}
\psfrag{x}{{\tiny $x$}}
\psfrag{t0}{{\tiny $t_0$}}
\psfrag{t1}{{\tiny $t_1$}}
\psfrag{h0}{{\tiny $h_0$}}
\psfrag{h0+}{{\tiny $h_0+\tau_0$}}
\psfrag{e}{{\tiny $\|e\|$}}
\psfrag{n}{{\tiny $\sigma \|x\|$}}
\psfrag{(a)}{{\tiny (a)}}
\psfrag{(b)}{{\tiny (b)}}
\includegraphics[width=0.48 \textwidth]{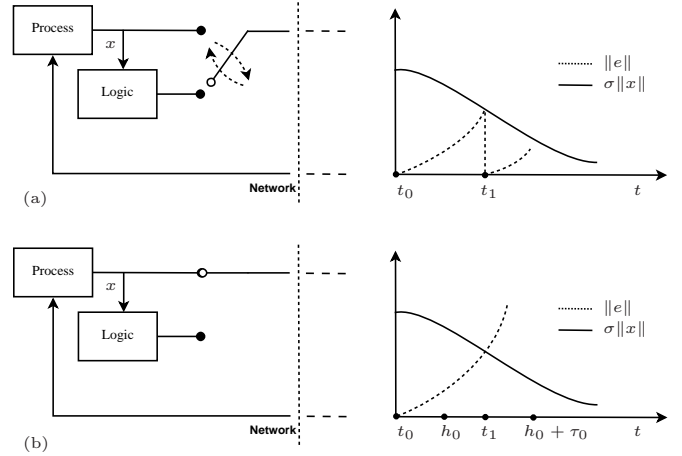} 
\linespread{1}\caption{Ideal mechanism for the fulfillment of (\ref{control_update_rule}): 
(a) absence of DoS;  
(b) presence of DoS.} \label{fig:noDoDvsDos}
\end{figure}

To fix the ideas, it is convenient to 
briefly comment on a possible implementation of condition (\ref{control_update_rule}),
referring the interested reader to Section \ref{sec:switch} for a thorough discussion and possible variations.
The simplest architecture one can think of for implementing (\ref{control_update_rule}) 
is that depicted in Figure \ref{fig:noDoDvsDos}(a).
The ``Logic'' block measures continuously the state $x$, computes the error signal $e$ and detects 
the instants $t_k$ at which (\ref{control_update_rule}) holds with the equality relation. At these instants, the logic 
samples the state and attempt to transmit it to the controller. In accordance with (\ref{error}), if the control 
update is successful then $e$ is reset to zero. 

If instead an acknowledgment is not received, 
then the logic infers that the packet is lost, that is, $t_k \in H_n$ 
for some $n\in \mathbb{N}$. Under such circumstances,
the logic turns to a different operating mode by 
continuously attempting to update the control action,
as depicted in Figure \ref{fig:noDoDvsDos}(b).
In this way, at time $h_{n}+\tau_n$ when communication is restored, 
the logic is able to transmit immediately the sampled measurement so that 
 (\ref{control_update_rule}) is enforced.

In the following subsection, for simplicity of exposition,
we assume that this is indeed the case.
In practice, when implementing  (\ref{control_update_rule}) on a digital platform, due to the finite sampling rate, a time interval will necessarily elapse from the time $h_n+\tau_n$ at which DoS is over, 
to the time at which the logic successfully samples and transmits. 
As anticipated, this case will be addressed in full details
in Section \ref{sec:switch}.

\subsection{Stability analysis}

We now study the trajectories of the closed-loop system 
composed of (\ref{system}) and (\ref{actual_control})
with control update law (\ref{control_update_rule}).
An alternative approach to stability analysis, based on Lyapunov functions, is discussed
in Appendix A.

Observe first that $\Phi$ is a stability matrix by hypothesis. 
Then there exist $\mu \in \mathbb R_{\geq 1}$ and $\lambda \in \mathbb R_{> 0}$ 
such that $\|e^{\Phi t}\| \leq \mu e^{-\lambda t}$ for all $t \in \mathbb R_{\geq0}$,
where $\mu$ and $\lambda$ can be easily computed using algebraic matrix theory.
This, in turns, implies
\begin{eqnarray} \label{control_system_bound}
\|x(t)\| \, \leq \, \omega_1 e^{-\lambda t}  &+&  \int_{\Theta(t)} \omega_2 \,  e^{-\lambda (t-s)} \|e(s)\| ds \nonumber \\
&+&  \int_{\Xi(t)} \omega_2 \, e^{-\lambda (t-s)} \|e(s)\| ds
\end{eqnarray}
having defined $\Theta(t):=[0,t)\backslash \Xi(t)$, $\omega_1 := \mu \|x(0)\|$ and $\omega_2 := \mu \|BK\|$
where, given a matrix $M$, $\|M\|$ denotes its spectral norm. 
We now evaluate the integral terms of the above formula 
separately.

Consider first the interval $\Theta(t)$, over which (\ref{control_update_rule}) holds
by construction. 
The corresponding integral term can be therefore 
upper bounded as 
\begin{eqnarray} \label{integral_term_1}
\int_{\Theta(t)} \omega_2 \,  e^{-\lambda (t-s)} \|e(s)\| ds \, \leq \, 
\int_{0}^t \omega_3 \,  e^{-\lambda (t-s)} \|x(s)\| ds
\end{eqnarray}
where $\omega_3:=\sigma \omega_2$.

Consider next $\Xi(t)$. In this case, some intermediate 
steps are needed. 
To begin with, we
avail ourselves of the following lemma.
\begin{lem} \label{lem:tech}
Consider the system $\Sigma$ composed of 
(\ref{system}) under a state-feedback control as in (\ref{actual_control})
with  control update rule (\ref{control_update_rule}). Then
\begin{eqnarray} \label{error_on_Hn_bound}
\| x(t_{k(h_n)})\| \, \leq \,  (1+\sigma) \|x(h_n)\|
\end{eqnarray}
for all $n \in \mathbb N$. \hfill $\Box$
\end{lem}

Consider the $n$-th DoS interval $H_n$.
Over each $H_n$, the process dynamics are 
governed by
\begin{eqnarray} \label{process_dynamics_on_Hn}
\dot x(t) \, = \, A x(t) + BK x(t_{k(h_n)}) 
\end{eqnarray}
In addition, there exist $\theta \in \mathbb R_{\geq 1}$ and $\rho \in \mathbb R_{\geq 0}$ 
such that $\|e^{A t}\| \leq \theta e^{\rho t}$ for all $t \in \mathbb R_{\geq0}$
\footnote{Alternatively, to possibly achieve tighter bounds, 
one can make use of the logarithmic norm of a matrix $M$,
which is defined as
\begin{eqnarray} \label{log_norm}
\rho_M \, := \, \max \left\{ \rho| \, \, \rho \in \textrm{spectrum} \left\{ (M+M^\top)/2 \right\} \right\}
\end{eqnarray}
where $M^\top$ is the transpose of $M$. In such a case, one has the bound
$\|e^{A t}\| \leq e^{\rho_A t}$ for all $t \in \mathbb R_{\geq0}$ \citep{Strom}.}.
Using (\ref{error_on_Hn_bound}) and (\ref{process_dynamics_on_Hn}), it is then
straightforward to verify that, over each DoS interval, 
the solution to the closed-loop dynamics 
can be upper bounded as 
\begin{eqnarray} \label{process_dynamics_on_Hn_bound}
\|x(t)\| \, &\leq& \,  \theta e^{\rho (t-h_n)} \|x(h_n)\| + \nonumber\\
&& 
\hspace{1.8cm}
\int_{h_n}^{t} \theta e^{\rho (t-\zeta)} \|BK\|  \|x(t_{k(h_n)}) \| d\zeta \nonumber \\
&\leq& \, \theta e^{\rho (t-h_n)} \|x(h_n)\| + \int_{h_n}^{t} \theta_1 e^{\rho (t-\zeta)} \|x(h_n)\| d\zeta \nonumber \\
&\leq& \, \theta_2 e^{\rho (t-h_n)}  \|x(h_n)\| 
\end{eqnarray}
having defined $\theta_1 =: \theta (1+\sigma) \|BK\|$ and $\theta_2 =: \theta + \theta_1/\rho$.
Recalling the definition of $\Xi(t)$, let 
\begin{eqnarray} \label{integral_term_2_tau}
\tau_n (t) \, := \min \{\tau_{n},t-h_{n}\} 
\end{eqnarray}
denote the length of the last DoS interval up to time $t$. 
In (\ref{control_system_bound}), the contribution to the evolution  of the state by each DoS interval can be upper bounded as follows
\begin{eqnarray} \label{integral_term_2_preliminary}
&& \int_{h_n}^{h_n+\tau_n(t)}  \omega_2 \, e^{-\lambda (t-s)} \|e(s)\| ds   \nonumber \\
&&
= \,  \int_{h_n}^{h_n+\tau_n(t)} \omega_2 \, e^{-\lambda (t-s)}  \| x(t_{k(h_n)})- x(s)\| ds \nonumber \\
&&
\leq \,  \int_{h_n}^{h_n+\tau_n(t)} \omega_2 \, e^{-\lambda (t-s)} \left[ \| x(t_{k(h_n)})\| + \|x(s)\| \right] ds \nonumber \\
&&\leq \, \int_{h_n}^{h_n+\tau_n(t)}  \omega_4 \, e^{-\lambda (t-s)} \, e^{\rho (s-h_n)} \,  \|x(h_n)\| ds \nonumber \\ \nonumber \\
&&= \,  \omega_*(\rho) \, e^{-\lambda (t-h_n)}  \left[  \, e^{(\lambda + \rho) \tau_n(t)}-1 \right] \|x(h_n)\|
\end{eqnarray}
for all $n \in \mathbb N$ with $n\leq n(t)$, where 
the second inequality follows from
(\ref{error_on_Hn_bound}) and (\ref{process_dynamics_on_Hn_bound})
with $\omega_4 =: \omega_2 (1+\sigma)+\omega_2 \theta_2$;
the last equality holds by letting $\omega_*(\rho)=:\omega_4/(\lambda+\rho)$.

Notice that by increasing $\rho$ if necessary one can always 
assume that  (\ref{integral_term_2_preliminary}) holds with $\omega_*(\rho) \leq 1$.
Specifically, let $\rho$ be any positive scalar such that  $\|e^{A t}\| \leq \theta e^{\rho t}$
for all $t \in \mathbb R_{\geq0}$ with $\theta \in \mathbb R_{\geq 1}$. Then, by letting
\begin{eqnarray} \label{integral_term_2_constants}
&& \rho_* \, := \, \inf \left\{ \zeta \in \mathbb R_{\geq\rho}  \, | \, \, \omega_*(\zeta) \leq 1 \right\}   \\ \nonumber \\
&& \delta_n (t) \, := \,  e^{(\lambda + \rho_*) \tau_n(t)}-1
\end{eqnarray}
one can always rewrite (\ref{integral_term_2_preliminary}) as
\begin{eqnarray} \label{integral_term_2}
\int_{h_n}^{h_n+\tau_n(t)} \omega_2 \, e^{-\lambda (t-s)} \|e(s)\| ds   \, \leq \, 
\delta_n(t) e^{-\lambda (t-h_n)} \|x(h_n)\| \nonumber \\
\end{eqnarray}
Hence, the last integral term of (\ref{control_system_bound}) 
can be finally upper bounded as 
 \begin{eqnarray} \label{integral_term_2}
&& \int_{\Xi(t)} \omega_2 \,  e^{-\lambda (t-s)} \|e(s)\| ds \, \leq \,  \sum_{n=0}^{n(t)} \, \delta_n(t) e^{-\lambda (t-h_n)} \|x(h_n)\| \nonumber \\ 
\end{eqnarray}
Combining (\ref{control_system_bound}), (\ref{integral_term_1})
and (\ref{integral_term_2}), the following results can be established.

\begin{theorem} \label{thm:gb}
Consider the system $\Sigma$ composed of 
(\ref{system}) under a state-feedback control as in (\ref{actual_control}).
Let $\Phi=A+BK$, with $\mu \in \mathbb R_{\geq 1}$ and $\lambda \in \mathbb R_{> 0}$ 
positive constants satisfying $\|e^{\Phi t}\| \leq \mu e^{-\lambda t}$ for all $t \in \mathbb R_{\geq0}$. Let
the control update rule satisfy (\ref{control_update_rule}) with 
\begin{equation} \label{control_update_rule_constraint_gb}
\lambda  - \sigma \mu \|BK\| >0
\end{equation}
Then, $\Sigma$ is GES for any DoS sequence $\{h_n\}$
satisfying Assumption \ref{ass:DoS_slow} with
\begin{equation} \label{tau_gb}
\tau \, > \,  \frac{\lambda + \rho_*}{\lambda  - \sigma \mu \|BK\|} 
\end{equation}
where $\rho_*$ is as in (\ref{integral_term_2_constants}).
In particular, under the stated conditions, (\ref{GES}) holds true with constants
$\alpha = \mu e^{\kappa(\lambda + \rho_*)}$ and $\beta = \lambda  - \sigma \mu \|BK\| - (\lambda + \rho_*)/\tau$. \hfill $\Box$
\end{theorem}

\begin{remark}  \label{rem:thm:gb}
The constraint (\ref{control_update_rule_constraint_gb}) must be satisfied
even in the absence of DoS. It reflects the fact that, even when 
communication is always possible, in order to achieve stability,
the control action must be updated frequently enough.
On the other hand, (\ref{tau_gb}) imposes constraints on the 
admissible DoS signals. In this respect, notice that
in contrast with hybrid system analysis where $\tau$ is allowed
to take on values less than one, here $\tau$ must always be greater 
than one. Such a constraint is 
consistent with intuition, reflecting the fact that, to achieve stability,
the total length of DoS intervals must be a suitable \emph{fraction}
of the time (in fact, $|\Xi(t)| \leq t/\tau$ when $\kappa=0$).
\hfill $\Box$
\end{remark}

\begin{remark}  \label{rem:thm:gb_2}
At the expense of possibly larger overshoots and more frequent control updates,
one can always design the control system so as to tolerate any
DoS signal not exceeding a prescribed fraction of time. Specifically,
for any given $\bar \tau >1$, one can always design $K$ yielding 
a nominal decaying rate $\lambda$ large enough for the 
right hand side of (\ref{tau_gb}) to be strictly less than $\bar \tau$.
While this will possibly result in a large value of $\mu \|BK\| $, 
condition (\ref{control_update_rule_constraint_gb}) can be 
still satisfied by properly selecting $\sigma$.
\hfill $\Box$
\end{remark}

\section{Implementation and resilient control logics}\label{sec:switch}

The analysis of Section 3 hinges upon the fulfillment of 
condition (\ref{control_update_rule}). 
Such a condition cannot be directly implemented on digital platforms
in that, in order to be fulfilled, it would require continuous
transmission attempts upon DoS detection, \emph{i.e.}
an \emph{infinite} sampling rate. 
Motivated by this, we first discuss how  
Theorem \ref{thm:gb} can be generalized
so as to possibly account for 
finite sampling rate constraints. 
Building upon Theorem \ref{thm:gb},
we finally consider a number of implementation possibilities
that can be used to trade-off performance vs. communication
resources within the proposed framework. 

\subsection{Stability under finite sampling rate} 

We first consider the following definition.

\begin{definition} \label{def:FSR}
A control update sequence $\{t_k\}$ 
is said to occur at a \emph{finite sampling rate} if
there exist an $\varepsilon \in \mathbb R_{>0}$
such that 
\begin{eqnarray} \label{minimal_IET}
 \Delta_k \, :=\,  t_{k+1} -t_k \, \geq \, \varepsilon 
\end{eqnarray}
for all $k \in \mathbb N$.
\hfill $\Box$
\end{definition}

Consider now a control update sequence $\{t_k\}$ along with a DoS sequence $\{h_n\}$,
and let
\begin{eqnarray} \label{}
\mathbb S_n \, := \,  \left\{ k \in \mathbb N \, |\,\, t_k \in H_n \right\}
\end{eqnarray} 
denote the set of integers associated with an attempt 
to update the control action during $H_n$. Accordingly,
by defining
\begin{eqnarray} \label{}
\Delta_{\mathbb S_n} \, := \, \sup_{k \in \mathbb S_n}  \Delta_k
\end{eqnarray} 
then
\begin{eqnarray} \label{DoS_interval_plus_FSR}
\bar H_n := [h_n,\, h_n + \tau_n + \Delta_{\mathbb S_n}[
\end{eqnarray}  
will provide an upper bound on the $n$-th time interval over which 
the control action is not updated, while
\begin{eqnarray}  \label{DoS_intervals_union_digital}
\bar \Xi(t) \, &:=& \, \left\{ \, \bigcup_{n=0}^{n(t)-1} \bar H_n \; \right\} \, \bigcup \nonumber \\ \nonumber \\
&&  \, \left[h_{n(t)}, \min \{h_{n(t)}+\tau_{n(t)}+\Delta_{\mathbb S_{n(t)}}; \,t\} \right] 
\end{eqnarray}
will provide an upper bound on the total interval up to the current time over which 
the control action is not updated. Equation (\ref{DoS_interval_plus_FSR})
essentially models the additional delay in the control update that 
may arise under finite sampling rate. In fact, 
under (\ref{minimal_IET}), $\Delta_{\mathbb S_n}$
will be greater than or equal to $\varepsilon$ so that $|\bar H_n|$
will be strictly greater than $|H_n|$. Notice that  (\ref{DoS_interval_plus_FSR}) 
is non-conservative in the sense $\bar H_n$ may be exactly equivalent to the $n$-th time 
interval over which the control action is not updated.
One may in fact have situations where a control update is requested 
just before the time $h_n + \tau_n$ at which the $n$-th DoS interval is over
and the next sampling time is scheduled at $h_n + \tau_n + \Delta_{\mathbb S_n}$.
Such a case cannot be ruled out being $h_n$ and $\tau_n$ unknown.

With this definition in place, the following result can be stated
which extends the conclusions of Theorem \ref{thm:gb} to
control update sequences possibly occurring at a finite sampling rate.
To avoid confusion, it is worth pointing out that the result which follows 
is only concerned with the effects of lower bounding $\{\Delta_k\}$ upon communication loss.
Logics satisfying the conditions of Theorem \ref{thm:gb_2} with explicit 
lower and upper bounds for $\Delta_*$
are discussed in the next subsection.

\begin{theorem} \label{thm:gb_2}
Consider the system $\Sigma$ composed of 
(\ref{system}) under a state-feedback control as in (\ref{actual_control}).
Let $\Phi=A+BK$, with $\mu \in \mathbb R_{\geq 1}$ and $\lambda \in \mathbb R_{> 0}$ 
positive constants satisfying $\|e^{\Phi t}\| \leq \mu e^{-\lambda t}$ for all $t \in \mathbb R_{\geq0}$. Let
the control update rule satisfy (\ref{control_update_rule}) with $\Xi(t)$ replaced by $\bar \Xi(t)$,
where $\sigma$ is as in (\ref{control_update_rule_constraint_gb}).
Then, $\Sigma$ is GES for any DoS sequence $\{h_n\}$
satisfying Assumption \ref{ass:DoS_slow} with
\begin{equation} \label{tau_gb}
\tau \, > \,  \left( \, \frac{\lambda + \rho_*}{\lambda  - \sigma \mu \|BK\|} \, \right) \left(\, 1+ \frac{\Delta_*}{\tau_*} \,\right)
\end{equation}
where
\begin{equation} \label{Delta_*}
\Delta_* \, := \, \sup_{n \in \mathbb N}  \Delta_{\mathbb S_n}
\end{equation}
and 
\begin{equation} \label{tau_*}
\tau_* \, := \, \inf_{n \in \mathbb N}  \tau_{n}
\end{equation}
In particular, under the stated conditions, (\ref{GES}) holds true with constants
$\alpha = \mu e^{(\lambda+\rho_*)(1+\Delta_*/\tau_*)\kappa}$ and 
$\beta = \lambda  - \sigma \mu \|BK\| - (\lambda + \rho_*)(1+\Delta_*/\tau_*)/\tau$. \hfill $\Box$
\end{theorem}

\begin{figure}[tb]
\psfrag{x}{{\tiny $x$}}
\psfrag{t}{{\tiny $t$}}
\psfrag{t1}{{\tiny $t_k$}}
\psfrag{t2}{{\tiny $t_{k+1}$}}
\psfrag{t3}{{\tiny $t_{k+2}$}}
\psfrag{h0}{{\tiny $h_n$}}
\psfrag{h1}{{\tiny $h_{n+1}$}}
\psfrag{e}{{\tiny $\|e\|$}}
\psfrag{n}{{\tiny $\sigma \|x\|$}}
\psfrag{(a)}{{\tiny (a)}}
\psfrag{(b)}{{\tiny (b)}}
\includegraphics[width=0.45 \textwidth]{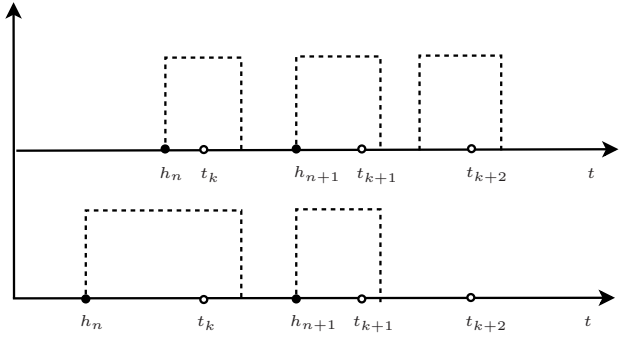} 
\linespread{1}\caption{Two DoS sequences (in dashed line) of equal total length.
The one at the top, composed of more intervals 
having smaller duration, denies more communications attempts
than the one at the bottom.} \label{fig:DoS_under_FSR}
\end{figure}

\begin{remark}  \label{rem:thm_2}
Theorem \ref{thm:gb_2} differs from Theorem \ref{thm:gb} not only 
because of $\Delta_*$ but also due to the presence of $\tau_*$.
This has a very intuitive explanation. In fact, in the ideal case considered in Theorem \ref{thm:gb}, 
$\Delta_*=0$ since a control update
can always occur as soon as DoS is over. 
Under finite sampling rate, each DoS interval will instead 
possibly introduce an additional delay in the control update.
Accordingly, given two DoS sequences 
of equal total length, the one composed of more intervals 
having smaller duration will be more critical for stability,
since it will potentially deny more communications attempts,
as depicted in Figure \ref{fig:DoS_under_FSR}. This can also
be seen from (\ref{tau_gb}): the smaller $\tau_*$, the larger the
value of $\tau$ required to achieve stability.
\hfill $\Box$
\end{remark}

\subsection{Implementation and resilient control logics}

The framework introduced with Theorem \ref{thm:gb_2}
is flexible enough so as to allow the designer to choose 
from several implementation options, some of which are described in the following.
Although these solutions originate from
fundamentally different approaches, they exhibit
the common feature of \emph{resilience}, by which
we mean not only to ensure a certain degree of 
robustness against DoS, but also the ability
to counteract it by changing the control update rule
upon communication loss.

\emph{Event/Time-driven logics}.
As discussed in Section 3,
the simplest architecture one can think of consists in
using a``Logic'' block that  measures continuously the state $x$, computes the error signal $e$ and detects 
the instants (events) at which 
\begin{eqnarray} \label{}
\|e(t)\| = \sigma  \|x(t)\|
\end{eqnarray}
At these instants, the logic 
samples the process state and attempt to transmit it to the controller. 
If an acknowledgment is not received, 
the logic turns to a different operating mode and 
attempts to update the control action periodically
\footnote{A periodic update is also enforced when $x(t_k)$=0.
This is because application of the second of (\ref{Tabuada_mixed}) for $x(t_k)$=0
would result in a continuous control update.}.

This is formalized in the next result.

\begin{proposition} \label{prop_1:gb}
Let $\Delta_1$ be a positive scalar less than or 
equal to $\Delta_2$, 
with $\Delta_2$ given by $\phi(\Delta_2) \,=\, \sigma$, the latter
being the unique solution at $\Delta_2$  of the generalized scalar Riccati equation
\begin{eqnarray} \label{minimal_IET_tab_eq}
\dot \phi(t) = \| \Phi \| + \left( \| \Phi \| + \| BK \| \right) \phi(t) + \| BK \| \phi^2(t) 
\end{eqnarray} 
initialized at $\phi(0)=0$, where $\Phi=A+BK$.   
Then, the control update rule
\begin{eqnarray}  \label{Tabuada_mixed}
t_{k+1} = \left\{ 
\begin{array}{ll} 
t_k + \Delta_1, & \quad \textrm{if} \,\, t_k \in H_{n(t)}  \\ & \quad  \textrm{or} \, x(t_k)=0 \\ \\
\inf \left\{ \, t \in \mathbb R_{>{t_k}}:\, \right. &  \\  
\qquad \left. \|e(t)\|  =  \sigma  \|x(t)\| \, \right\},  & \quad \textrm{otherwise}
\end{array}
\right. 
\end{eqnarray}
satisfies the conditions of Theorem \ref{thm:gb_2} with $\Delta_*=\Delta_1$
and $\Delta_k \geq \Delta_1$ for all $k \in \mathbb N$. \hfill $\Box$
 \end{proposition}

\begin{figure}[tb]
\psfrag{x}{{\tiny $x$}}
\psfrag{u}{{\tiny $u$}}
\psfrag{t0}{{\tiny $t_0$}}
\psfrag{t1}{{\tiny $t_1$}}
\psfrag{h0}{{\tiny $h_0$}}
\psfrag{h1}{{\tiny $h_1$}}
\psfrag{e}{{\tiny $\|e\|$}}
\psfrag{n}{{\tiny $\sigma \|x\|$}}
\psfrag{(a)}{{\tiny (a)}}
\psfrag{(b)}{{\tiny (b)}}
\includegraphics[width=0.45 \textwidth]{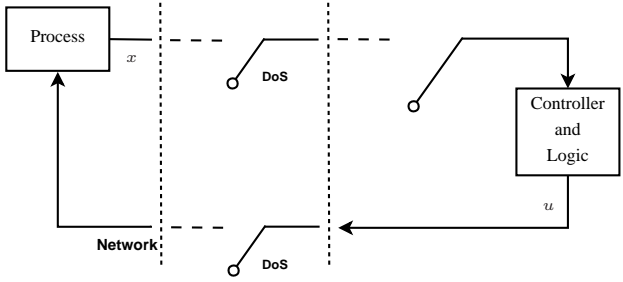} 
\linespread{1}\caption{Implementation for time-driven 
and self-triggering policies with the ``Logic'' block embedded in the control unit.} \label{fig:time_driven_logics}
\end{figure}

\emph{Purely time-driven logics}. The rationale 
behind (\ref{Tabuada_mixed}) is  that, upon DoS detection,
transmission is attempted at the sampling rate specified by $\Delta_1$, 
while, in the absence of DoS, 
less frequent control updates are allowed. 
This mechanism has the positive feature of potentially saving
communication resources but requires  
continuous process state monitoring. 
Unless dedicated hardware is available for this purpose,
it is convenient to search for purely time-driven logics
in which the ``Logic'' block is embedded in the control unit,
as depicted in Figure \ref{fig:time_driven_logics}.

One possible solution is captured in the next result, whose 
proof follows directly from Proposition \ref{prop_1:gb}.

\begin{proposition} \label{prop_2:gb}
Let $\Delta_1$ and $\Delta_2$ be positive scalars 
with $\Delta_1\leq \Delta_2$  and $\Delta_2$
as in Proposition \ref{prop_1:gb}. 
Then, the control update rule
\begin{eqnarray}  \label{Periodic_mixed}
t_{k+1} = \left\{ 
\begin{array}{ll} 
t_k + \Delta_1, & \qquad \qquad \qquad \textrm{if} \,\,  t_k 
\in H_{n(t)}   \\ \\
t_k + \Delta_2,  & \qquad \qquad \qquad \textrm{otherwise}
\end{array}
\right. 
\end{eqnarray}
satisfies the conditions of Theorem \ref{thm:gb_2} with $\Delta_*=\Delta_1$
and $\Delta_k \geq \Delta_1$ for all $k \in \mathbb N$. \hfill $\Box$
 \end{proposition}

\emph{Self-triggering logics}. As a final possibility, we note that
purely time-driven logics can be relaxed to more flexible aperiodic
implementations by letting $\Delta_k$ to take values 
based on the available data. Logics of this kind
are typically referred to as ``self-triggering'' in that
the next update instant is computed directly by the 
control unit. Let $t_1,t_2 \in \mathbb R_{\geq 0}$ with $t_2\geq t_1\geq0$ and define
\begin{eqnarray}  \label{}
\chi(t_2,t_1) := \left[ \, e^{\Phi(t_2-t_1)}  + \int_{t_1}^{t_2}  e^{\Phi(t_2-s)} BK ds \, \right] x(t_1)  
\end{eqnarray}
Thus $\chi(t_k,t_{k(t)})$ provides a prediction of $x(t_k)$
based on the last successful measurement $x(t_{k(t)})$.
Thus, one can set $\Delta_k$ depending on the 
magnitude of $\|\chi(t_k,t_{k(t)})\|$: the larger $\|\chi(t_k,t_{k(t)})\|$ the 
smaller $\Delta_k$ and viceversa, which corresponds to increasing the sampling rate as 
the distance of the process state from the origin gets larger.  

Then the next result holds, which 
follows again from Proposition \ref{prop_1:gb}
\footnote{A function $\varphi: \mathbb R_{\geq 0} \mapsto \mathbb R_{\geq 0}$, 
is said to be of class $\mathcal K$ if it is continuous, strictly increasing and
$\varphi(0)=0$.}. 

\begin{proposition} \label{prop_3:gb}
Let $\Delta_1$ and $\Delta_2$ be positive scalars 
with $\Delta_1\leq \Delta_2$  and $\Delta_2$
as in Proposition \ref{prop_1:gb}. Let $\varphi: \mathbb R_{\geq 0} \mapsto [0,\,1]$, 
be a class $\mathcal K$ function.
Then, the control update rule
\begin{eqnarray}  \label{self_triggering}
t_{k+1} =  t_k + \Delta_2 - (\Delta_2 - \Delta_1) \varphi \left( \left\| \chi(t_k,t_{k(t)}) \right\| \right)
\end{eqnarray}
satisfies the conditions of Theorem \ref{thm:gb_2} with $\Delta_*=\Delta_1$
and $\Delta_k \geq \Delta_1$ for all $k \in \mathbb N$.  \hfill $\Box$
 \end{proposition}

%

\section{Conclusions}

We have studied resilient control strategies for linear systems under DoS. 
We have shown that to conclude asymptotic stability, DoS signals must not 
be active for more than a certain percentage of time on the average. 
The resilient nature of the  proposed control strategy descends from its ability 
to adapt  the sampling rate to the state of the process and to the occurrence of DoS attacks.  
The results lend themselves to be extended in various directions.
We have not investigated the effect of possible limitations on the information, 
such as disturbances, quantization and delays, and leave the topic for future investigation.  
As additional future research topics, we envision the use of similar techniques 
to handle the problem in the presence of output feedback and for nonlinear systems. 
Regarding the latter extension, the alternative Lyapunov-based analysis of the problem  presented in Appendix A suits well our purpose.
One of the motivation to consider control problems over networks descends from problems of distributed coordination and control of large-scale systems. Investigating our approach to resilient control under DoS for event-based coordination problems such as those in \cite{CDP:PF:TAC13} represents another interesting research venue.

\appendix

\section*{Appendix}

\section{Lyapunov-based approach} \label{sub:proporties}

Lyapunov arguments provide an alternative analysis of the problem that is sometimes useful. 
Unless otherwise stated, the notation for this section is the same as in Section 3. 
Consider again the control system composed of 
(\ref{system}) under a state-feedback control as in (\ref{actual_control})
with control update rule (\ref{control_update_rule}). 
Given any positive definite matrix $Q=Q^\top \in \mathbb R^{n_x \times n_x}$,
let $P$ be the unique solution of the Lyapunov equation
\begin{eqnarray} \label{lyap}
\Phi^\top P + P \, \Phi + Q = 0
\end{eqnarray}
Then, by taking $V(x) = x^\top P x$ as a Lyapunov function, 
and computing it along the solution to (\ref{control_system}), 
it is immediate to see that 
\begin{eqnarray} \label{lyap_constants_0}
&& \alpha_1 \|x(t)\|^2 \, \leq \,  V(x(t)) \, \leq \, \alpha_2 \|x(t)\|^2   \\ \nonumber \\  \label{lyap_constants_1}
&& \dot V(x(t)) \, \leq \, -\gamma_1  \|x(t)\|^2 + \gamma_2 \|x(t)\|  \|e(t)\|
\end{eqnarray}
hold for all $t \in \mathbb R_{\geq0}$, with  
$\alpha_1$ and $\alpha_2$ equal to the smallest and largest 
eigenvalue of $P$, respectively, $\gamma_1$ equal to the smallest
eigenvalue of $Q$, and $\gamma_2:=\|K^\top B^\top P + PBK\|$
(\emph{cf.} \cite{Tabuada07}).

Consider first $\Theta(t)=[0,t)\backslash \Xi(t)$, over which (\ref{control_update_rule}) holds
by construction. In this case, simple calculations yield
\begin{eqnarray} \label{lyap_bound_1}
\dot V(x(t)) \, \leq \, -\omega_1  V(x(t)) 
\end{eqnarray}
where $\omega_1:= (\gamma_1  - \gamma_2 \sigma)/\alpha_2$.

Consider next $\Xi(t)$. In this case, as before, some intermediate 
steps are needed. Consider the $n$-th DoS interval $H_n$.
From Lemma \ref{lem:tech} we have
\begin{eqnarray} \label{lyap_error_bound_1}
\|e(t)\| \, &=& \, \|x(t_{k(h_n)}) - x(t)\| \nonumber \\ \nonumber \\
&\leq& \, (1+\sigma) \|x(h_n)\| + \|x(t)\| 
\end{eqnarray}
for all $t \in H_n$.

Thus, for all $t \in H_n$ such that $\|x(h_n)\| \leq \|x(t)\|$,
one has
\begin{eqnarray} \label{lyap_bound_2}
\dot V(x(t)) \, &\leq& \, -\gamma_1  \|x(t)\|^2 + \gamma_2 (2+\sigma) \|x(t)\|^2 \nonumber \\ \nonumber \\
&<& \, \omega_2 V(x(t))
\end{eqnarray}
where $\omega_2:= \gamma_2 (2+\sigma)/\alpha_1$. 
Conversely, for all $t \in H_n$ such that $\|x(h_n)\| > \|x(t)\|$,
one has
\begin{eqnarray} \label{lyap_bound_3}
\dot V(x(t)) \, < \, \omega_2 V(x(h_n))
\end{eqnarray}
Combining the last two inequalities with (\ref{lyap_bound_1}),
the following result can be established.

\begin{theorem} \label{thm:lyap}
Consider the system $\Sigma$ composed of 
(\ref{system}) under a state-feedback control as in (\ref{actual_control}). 
Given any positive definite matrix $Q=Q^\top \in \mathbb R^{n_x \times n_x}$,
let $P$ be the unique solution of the Lyapunov equation 
$\Phi^\top P + P \, \Phi + Q = 0$ with
$\Phi=A+BK$. Let $V(x) = x^\top P x$, and let
the control update parameter $\sigma$ in (\ref{control_update_rule}) be such that
\begin{equation} \label{control_update_rule_constraint_lyap}
\gamma_1  - \sigma \gamma_2 >0
\end{equation}
with $\gamma_1$ and $\gamma_2$ as in (\ref{lyap_constants_1}).
Then, $\Sigma$ is GES for any DoS sequence $\{h_n\}$
satisfying Assumption \ref{ass:DoS_slow} with
\begin{equation} \label{tau_lyap}
\tau \, > \,  \frac{\omega_1 + \omega_2}{\omega_1} 
\end{equation}
where $\omega_1= (\gamma_1  - \gamma_2 \sigma)/\alpha_2$ and $\omega_2= \gamma_2 (2+\sigma)/\alpha_1$,
and $\alpha_1$ and $\alpha_2$ as in (\ref{lyap_constants_0}).
In particular, under the stated conditions, (\ref{GES}) holds true with constants
$\alpha = \sqrt{e^{\kappa(\omega_1+\omega_2)} \alpha_2/\alpha_1}$ and $\beta = [\omega_1 -  (\omega_1+\omega_2)/\tau]/2$.
\end{theorem}

\emph{Proof}. 
For notational convenience, let $h_{-1}=\tau_{-1}:=0$.
From (\ref{lyap_bound_1}) we have
\begin{eqnarray} \label{lyap_bound_4}
V(x(t)) \, \leq \, e^{-\omega_1 (t-h_{n-1}-\tau_{n-1})}  V(x(h_{n-1}+\tau_{n-1})) 
\end{eqnarray}
for all $t \in [h_{n-1}+\tau_{n-1},h_{n}[$ with $n \in \mathbb N$.
In addition, (\ref{lyap_bound_2}) and  (\ref{lyap_bound_3}) imply
\begin{eqnarray} \label{lyap_bound_5}
V(x(t)) \, \leq \, e^{\omega_2 (t-h_n)}  V(x(h_n)) 
\end{eqnarray}
for all $t \in H_n$ with $n \in \mathbb N$.

Combining the last two expressions and recalling the definitions 
of $\Theta(t)$ and $\Xi(t)$, we get
\begin{eqnarray} \label{lyap_bound_6}
V(x(t)) \, &\leq& \, e^{-\omega_1 |\Theta(t)|} \, e^{\omega_2\,  |\Xi(t)|} \, V(x(0))  \nonumber \\ \nonumber \\
&=& \, e^{-\omega_1 t } \, e^{ (\omega_1+\omega_2)\,  |\Xi(t)|} \, V(x(0)) 
\nonumber \\ \nonumber \\
&=& \, e^{\kappa(\omega_1+\omega_2)} \, e^{-[\omega_1 -  (\omega_1+\omega_2)/\tau]\,t} \, V(x(0)) 
\end{eqnarray}
where the first equality follows since $ |\Theta(t)| = t -  |\Xi(t)|$. 
This establishes GES under the standing assumptions of the theorem. As for the computation of the constants,
it is sufficient to observe that (\ref{lyap_bound_6}), along with (\ref{lyap_constants_0}), implies  
\begin{eqnarray} \label{lyap_bound_7}
\|x(t)\|^2 \, &\leq& \, \frac{\alpha_2}{\alpha_1} e^{\kappa(\omega_1+\omega_2)} \, e^{-[\omega_1 -  (\omega_1+\omega_2)/\tau]\,t}  \, \|x(0)\|^2
\end{eqnarray}
which yields the desired result.
\hfill \qedp

\section{Proofs}\label{app.proofs}

\textbf{Proof of Lemma \ref{lem:tech}.} 
If $h_0=0$, then $x(t_{k(h_0)}) = 0$ by definition 
so that (\ref{error_on_Hn_bound}) is valid.
Recall next that
\begin{eqnarray} \label{error_on_Hn}
e(t) \, = \, x(t_{k(h_n)}) - x(t)
\end{eqnarray}
for all $t \in H_n$, $n \in \mathbb  N$, with $x(t_{k(h_n)})$ representing
the value of the process state at the last successful control update before $H_n$. 
Since (\ref{control_update_rule}) holds true for all $t \notin \Xi(t)$ 
and by continuity of $x(\cdot)$, one has 
$\|e(h_n)\| \leq \sigma \|x(h_n)\|$ when 
$h_0>0$ and for all $n \in \mathbb N_1$. 
Hence, 
\begin{eqnarray} 
\| x(t_{k(h_n)}) - x(h_n)\| \, \leq \,  \sigma \|x(h_n)\|
\end{eqnarray}
and (\ref{error_on_Hn_bound}) follows by applying the triangular inequality.
\hfill \qedp

The proof of Theorem  \ref{thm:gb} rests on  the following result, which is  a simplified variant of the
Gronwall type inequality for piecewise continuous functions considered in
\cite[Theorem 16.4]{Bainov}.

\begin{lem} \label{lem:gb}
Let $\{\ell_k\}$, $k \in \mathbb N$, be a fixed sequence 
satisfying $0 \leq \ell_0<\ell_1<\ell_2<\ldots$ and $\lim_{k \rightarrow \infty }\ell_k =\infty$.
Suppose that for $t \geq \ell_0$ we have
\begin{eqnarray} \label{gb_0}
\xi(t) \, \leq \, \omega_1(t) + \int_{\ell_0}^t \omega_2 \, \xi(s) ds+ \sum_{k \in \mathbb N_{1}:\, \ell_0<\ell_k<t} \delta_k(t) \, \xi(\ell_k) \nonumber \\
\end{eqnarray}
where $\xi: \mathbb R_{\geq 0} \mapsto \mathbb R_{>0}$ is continuous; 
$\omega_1 : \mathbb R_{\geq 0} \mapsto \mathbb R_{>0}$ is a nondecreasing function in $\mathbb R_{\geq 0}$;
$\omega_2 \in \mathbb R_{\geq 0}$ is a constant; and
$\delta_k(t) \in \mathbb R_{\geq 0}$ with $k \in \mathbb N_{1}$ are nondecreasing functions in $\mathbb R_{\geq 0}$.
Then
\begin{equation} \label{gb_1}
\xi(t) \, \leq \, \omega_1 \, e^{\omega_2 \,(t-\ell_0)} 
\, \prod_{k \in \mathbb N_{1}:\, \ell_0<\ell_k<t} (1+\delta_k(t) )  
\end{equation}
for all $t \geq \ell_0$. \hfill $\Box$
\end{lem}

When $\delta_k(t)=0$ for all $t \in \mathbb R_{\geq0}$ then 
(\ref{gb_0}) and (\ref{gb_1}) give rise to the standard Gronwall-Bellman inequality
\cite[Theorem 1.1]{Bainov}.

The idea is then to apply Lemma \ref{lem:gb} to the inequality obtained from the combination of (\ref{control_system_bound}), (\ref{integral_term_1})
and (\ref{integral_term_2}), with $\xi(\cdot)$ and $\{\ell_k\}$
changed into $\|x(\cdot)\|$ and $\{h_n\}$, respectively, using the latter
to model discontinuities in the process behavior caused by DoS.

\textbf{Proof of Theorem \ref{thm:gb}.} 
Define
\begin{eqnarray} \label{scaling_variable}
\xi(t) \, := \, e^{\lambda t} \|x(t)\|
\end{eqnarray}
Thus, using (\ref{integral_term_1})
and (\ref{integral_term_2}), one can rewrite (\ref{control_system_bound}) as
\begin{eqnarray} \label{control_system_bound_eta}
\xi(t) \, \leq \, \omega_1  +  \int_{0}^t \omega_3 \,  \xi(s) ds +  \sum_{n=0}^{n(t)} \, \delta_n(t) \xi(h_n)
\end{eqnarray}
Assume first $h_0>0$.
The conditions of Lemma \ref{lem:gb} are satisfied by letting 
$\ell_0=0$ and $\ell_{n+1}=h_n$ for all $n \in \mathbb N$, 
with $\delta_n(t)$ nondecreasing in $\mathbb R_{\geq 0}$ by construction.
Thus
\begin{eqnarray} \label{control_system_bound_eta_final}
\xi(t) \, &\leq& \, \omega_1 \, e^{\omega_3 \,t} 
\, \prod_{n=0}^{n(t)}  e^{(\lambda + \rho_*) \tau_n(t)} \nonumber \\ \nonumber \\
&=& \, \mu\,\xi(0) \, e^{\sigma \mu \|BK\|  \,t} 
\, e^{(\lambda + \rho_*) |\Xi(t)|}
\end{eqnarray}
for all $t \in \mathbb R_{\geq 0}$. Replacing in the inequality above
the expressions of the various variables and parameters, one obtains 
\begin{eqnarray} \label{}
\|\xi(t)\| \, &\leq& \, \mu \|x(0)\| 
\, e^{(\lambda + \rho_*)\kappa} \, e^{-(\lambda - \sigma \|BK\| - \frac{\lambda + \rho_*}{\tau} )t}
\end{eqnarray}
This yields the desired result.
Assume next $h_0=0$. Recalling that $\omega_1=\mu\,\xi(0)$,
one can rewrite (\ref{control_system_bound_eta}) as
\begin{eqnarray} \label{control_system_bound_eta_2}
\xi(t) \, \leq \, ( \mu+\delta_0(t) )\xi(0)  +  \int_{0}^t \omega_3 \,  \xi(s) ds +  \sum_{n=1}^{n(t)} \, \delta_n(t) \xi(h_n) \nonumber \\
\end{eqnarray}
Since $\mu \in \mathbb R_{\geq 1}$, we have
$\mu+\delta_0(t) \leq \mu e^{(\lambda + \rho_*) \tau_0(t) }$,
with $\tau_0(t)$ nondecreasing in $\mathbb R_{\geq 0}$ by construction.
Thus, by applying Lemma \ref{lem:gb} with
$\ell_{n}=h_n$ for all $n \in \mathbb N_1$, we obtain again (\ref{control_system_bound_eta_final}).
\hfill \qedp

\textbf{Proof of Theorem \ref{thm:gb_2}}. 
By definition of $\bar \Xi(t)$, we get
\begin{eqnarray} \label{}
|\bar \Xi(t)| \,&=& \, \sum_{n=0}^{n(t)-1} (\tau_n +  \Delta_{\mathbb S_n}) \nonumber \\ \nonumber \\
&+& \, \min \{ \tau_{n(t)} +  \Delta_{\mathbb S_{n(t)}}; \,t - h_{n(t)} \} 
\end{eqnarray} 
In addition,
\begin{eqnarray} \label{}
|\Xi(t)| \,= \, \sum_{n=0}^{n(t)-1} \tau_n + \min \{ \tau_{n(t)}; \,t - h_{n(t)} \}  
\end{eqnarray} 
If $\tau_{n(t)} > t - h_{n(t)}$, then $|\Xi(t)| \geq n(t) \, \tau_*$.
In turn, 
this yields
\begin{eqnarray} \label{}
|\bar \Xi(t)| \,&=& \, |\Xi(t)| +  \sum_{n=0}^{n(t)-1}  \Delta_{\mathbb S_n}
\nonumber \\ \nonumber \\
&\leq& \,  |\Xi(t)| +  n(t) \Delta_* \, \leq \, |\Xi(t)| \left( 1+ \frac{\Delta_*}{\tau_*} \, \right) 
\end{eqnarray} 
If instead $\tau_{n(t)} \leq t - h_{n(t)}$, then $|\Xi(t)| \geq (n(t)+1) \, \tau_*$.
This yields
\begin{eqnarray} \label{}
|\bar \Xi(t)| \,&\leq& \, |\Xi(t)| +  \sum_{n=0}^{n(t)}  \Delta_{\mathbb S_n} 
\nonumber \\ \nonumber \\
&\leq& \,  |\Xi(t)| +  (n(t)+1) \Delta_* \, \leq \, |\Xi(t)| \left( 1+ \frac{\Delta_*}{\tau_*} \, \right) \nonumber \\
\end{eqnarray} 
Thus, under Assumption \ref{ass:DoS_slow},
\begin{eqnarray} \label{}
|\bar \Xi(t)| \,&\leq& \,  \left( \kappa+ \frac{t}{\tau} \, \right)  \left( 1+ \frac{\Delta_*}{\tau_*} \, \right) 
\end{eqnarray} 
and the conclusion is that the proof of Theorem \ref{thm:gb} carries over to Theorem \ref{thm:gb_2}
with $\Xi(t)$ and $\Theta(t)$ replaced by $\bar \Xi(t)$ and 
$\bar \Theta(t):=[0,t)\backslash \bar \Xi(t)$, respectively.
\hfill \qedp

\emph{Proof of Proposition \ref{prop_1:gb}}. 
As shown in \cite{Tabuada07}, in the absence
of DoS, the control update rule
\begin{eqnarray}
t_{k+1} = \inf \left\{ \, t \in \mathbb R_{>{t_k}}:\,  \|e(t)\| = \sigma  \|x(t)\| \, \right\} \nonumber
\end{eqnarray}
ensures that 
\begin{eqnarray} \label{proof_prop_1}
\|e(t)\| \leq \sigma  \|x(t)\|
\end{eqnarray}
for all $t \in [t_k,\,t_{k+1}[$
with $\Delta_k \geq \Delta_2$. 
This also implies that, in the absence
of DoS, any periodic control update rule 
$t_{k+1} =t_k + \Delta_1$ with $\Delta_1\leq\Delta_2$
will still satisfy (\ref{proof_prop_1}) for all $t \in [t_k,\,t_{k+1}[$.
With this in mind, consider a DoS interval $H_n$. 
Then, two possible cases arise: 
if $h_{n+1} \leq h_n +\tau_n + \Delta_1$, which 
means that $\bar H_n$ and $\bar H_{n+1}$ overlap
each other, there is nothing to prove. If instead
$h_{n+1} > h_n +\tau_n + \Delta_1$, 
a successful control update will necessarily occur  
in the interval $[h_n +\tau_n,\,h_n +\tau_n + \Delta_1[$.
Since $\Delta_1\leq\Delta_2$, we conclude that 
(\ref{Tabuada_mixed}) satisfies the conditions of 
Theorem \ref{thm:gb_2} with $\Delta_* = \Delta_1$ by construction.
\hfill \qedp

\bibliographystyle{plainnat} 

\bibliography{Automatica_PAPER_BIBLIO}

\end{document}